\documentclass[preprint]{ptephy_v1}

\usepackage{amsfonts}

\usepackage{boites}
\usepackage{amsmath}
\usepackage{amsthm}
\usepackage{mathtools}
\usepackage{mathrsfs}
\usepackage{color}
\usepackage{braket}
\usepackage{url}
\usepackage{bm}
\usepackage[colorlinks=true,linkcolor=blue,citecolor=blue]{hyperref}
\usepackage{lineno}
\preprintnumber{NITEP 159}

\usepackage{comment}

\usepackage[normalem]{ulem}  

\newcommand{\br}{\boldsymbol{r}}

\begin{document}

\title{Electron wave functions in beta-decay formulas revisited (II): Completion including recoil-order and induced currents}

\author[1,2,3,4]{Wataru Horiuchi}
\affil[1]{Department of Physics, Osaka Metropolitan University, Osaka 558-8585, Japan}
\affil[2]{Nambu Yoichiro Institute of Theoretical and Experimental Physics (NITEP), 
Osaka Metropolitan University, Osaka 558-8585, Japan}
\affil[3]{RIKEN Nishina Center, Wako, Saitama 351-0198, Japan}
\affil[4]{Department of Physics, Hokkaido University, Sapporo 060-0810, Japan}

\author[5]{Toru Sato}
\affil[5]{Research Center for Nuclear Physics, Osaka University, Ibaraki, Osaka 567-0047, Japan} 
     
\author[6]{Yuichi Uesaka}
\affil[6]{Faculty of Science and Engineering, Kyushu Sangyo University, Fukuoka 813-8503, Japan}

\author[5,3]{Kenichi Yoshida}



\begin{abstract}
We present complete formulas of 
the allowed and first-forbidden transitions of 
the nuclear beta decay taking into account 
the recoil-order and induced currents 
up to the next-to-leading order (NLO). 
The longitudinal part of the vector current 
is cleared away by the use of the conservation 
of the vector current for the multipole operators of the 
natural-parity transitions, which 
makes the effect of the meson 
exchange current for the vector current as small as possible.
The formula is transparent enough to be applied 
to various beta-decay processes. 
As a numerical demonstration, we apply our formulas 
to the beta decay of a neutron-rich nucleus $^{160}$Sn.
We find that the NLO corrections amount to 10--20\% 
of the total decay rate, whereas the induced currents 
alter the rate at most 5\%.

\end{abstract}

\maketitle
\allowdisplaybreaks[1]

\section{Introduction}

Nuclear beta decay has attracted renewed interest.
The astrophysical rapid-neutron-capture process, the r-process nucleosynthesis, 
is believed to produce more than half of heavier elements than iron. 
The beta decay is a predominant process in the freeze-out phase~\cite{Mumpower:2015ova}, 
and thus plays a decisive role in forming the abundance pattern. 
The detection of gravitational waves from a binary neutron-star merger 
and associated electromagnetic waves from a kilonova/macronova has 
provided an evidence of the r-process site~\cite{abb17a,abb17b}.
In such an exotic environment, heavy neutron-rich nuclei 
even close to the dripline are involved in the r-process.

An interest in the nuclear beta decay with a precise description also lies in neutrino physics,
 such as the reactor neutrino anomaly~\cite{mention2011,hayes2014}
 and the search for the beyond-standard-model (BSM) physics~\cite{Glick_Magid_2022}.
The long-standing issue of nuclear physics,
the effective (weak) axial vector coupling constant ($g_A^{\rm eff}$),
might be resolved in the electron spectrum of the higher-order-forbidden transitions~\cite{kostensalo2017}. 

 In our previous paper~\cite{horiuchi2021electron} [referred to as paper (I)],
formulas of the beta-decay rate in a convenient and practical form were proposed.
The nuclear beta decay was formulated using
the effective multipole operators~\cite{Koshigiri:1979zd,morita1960theory}.
The analytic forms of the electron wave function in the
next-to-leading order (NLO) of the integral equation of the electron
Dirac equation and the formulas of the effective nuclear multipole operators
were presented. It was demonstrated that
the NLO formula reproduces well the `exact' decay rate 
of the allowed Gamow--Teller and first-forbidden spin-dipole transitions
for heavy nuclei with large $Z$, where the frequently used formula\cite{sch66}
corresponding to a leading order formula of the systematic
expansion in terms of nuclear radius, the Coulomb potential, the energy and the mass
of electron\cite{Behrens:1971rbq}, possibly overestimates the decay rates.

The formula in paper (I)
is expressed  in terms of the nuclear current density
that is general enough to incorporate any models of the nuclear current
such as many-body nuclear current in addition to the one-body current,
non-relativistic and relativistic description of nuclear current.
However, the explicit formula in the impulse approximation is given
only for the leading $\bm{\sigma}\tau$ operator  of the axial
vector current. It is certainly necessary and valuable to
update the formula of paper (I) by  including the velocity-dependent terms
such as $A_0 \sim \bm{\sigma}\cdot \bm{p}/M \tau$,
$\bm{V} \sim \bm{p}/M \tau$ and the induced currents.

In this paper,  we give the formulas of the nuclear multipole operators
using the non-relativistic impulse approximation of the nuclear current.
The velocity-dependent terms of the time component of the axial-vector current, 
the space component of the vector current and the induced terms, i.e., the weak-magnetism term 
and the induced pseudo-scalar term, are included.
These nuclear currents play an important role in the first-forbidden transitions.
For the $0^-$ transition, for example, 
the time component of the axial-vector current competes
with the spin-dipole operator~\cite{towner1986}.
The induced pseudo-scalar term
would play  a  minor role in the nuclear beta decay. However, we keep it
because the formulas can be used for the neutrino reactions including muon and tau leptons,
while the so-called second class currents, 
the G-parity irregular terms~\cite{weinberg58,wilkinson1977second}, are not included.

In neutrino reactions and beta decay in the plane-wave approximation of lepton wave functions, a
longitudinal part of the vector current is eliminated using the conservation of
current, and further the LO term of the transverse electric current
can be written in terms of the Coulomb multipole
operator~\cite{walecka2012semileptonic,Nakamura:2000vp,hayes2018}.
In the beta decay or muon capture, because of the complication due to the Coulomb effect,
this is done in the leading order of the long-wave-length 
approximation. In the muon capture, this was discussed in connection with the
extended Siegert's theorem~\cite{fujii-fujita-morita-64}.
In this paper, we rewrite the formula of paper (I) using the Fourier--Bessel
transformation and the current conservation, which makes the effect
of the meson-exchange current for the vector current as small as possible
for the allowed and first-forbidden transitions.
This work provides complete formulas of the allowed and first-forbidden transitions
of the nuclear beta decay up to NLO within the non-relativistic impulse approximation of the nuclear current.

In section \ref{sec:multipole_expansion}, the multipole operators for the parity $P=(-1)^J$ and $(-1)^{J+1}$
are given. The longitudinal part of the vector current
is rewritten in terms of the time component of the vector current
with the help of the current conservation.
In section \ref{sec:impilse_approx}, a formula of the multipole operators in the impulse approximation is given. 
As a numerical demonstration of the usefulness of our method, the formula is applied to the beta decay of $^{160}$Sn
in section \ref{sec:applications} together with the formula written in terms of
nuclear transition densities.
The transition densities are obtained by a nuclear energy-density-functional method.
A summary is given in section \ref{sec:summary}.

\section{Multipole expansion of the effective Hamiltonian}
\label{sec:multipole_expansion}

The  effective Hamiltonian for the beta ($\beta^\mp$)
decay is  given by
\begin{align}
    H_\mathrm{eff}=\frac{G_FV_{ud}}{\sqrt{2}}\int d\bm{x}\left[\overline{e}(\bm{x})\gamma^\mu(1-\gamma_5)\nu_e(\bm{x})J_\mu(\bm{x})+{\rm H.c.}\right],
\end{align}
where $G_F=1.166\times 10^{-5}$\,GeV$^{-2}$ is the Fermi coupling constant and $V_{ud}=0.9737$ is the up-down element of the Cabibbo--Kobayashi--Maskawa matrix.
This Hamiltonian consists of the electron field $e$, the neutrino field $\nu$, and the hadronic current
\begin{align}
    J^\mu(\bm{x})={\cal V}^\mu(\bm{x})-{\cal A}^\mu(\bm{x}),
\end{align}
with the vector [${\cal V}^\mu = (V^0,\bm{V})$] and axial-vector
[${\cal A}^\mu = (A^0,\bm{A})$] currents.
Using the multipole operator $\Xi_{JL}(\kappa_e,\kappa_\nu)$
constructed from partial waves of electron and neutrino specified in terms of
$\kappa_e$ and $\kappa_\nu$~\cite{horiuchi2021electron},
one obtains the rate of the $\beta^\mp$ decay 
from the initial ($i$) to final ($f$) nuclear states, $i\to f+e^\mp+\overline{\nu}_e(\nu_e)$, as
\begin{align}
    \Gamma=\frac{\left(G_FV_{ud}\right)^2}{\pi^2}\int_{m_e}^{E_0}dE_e\, p_eE_e\left(E_0-E_e\right)^2\sum_{J,\kappa_e,\kappa_\nu}\frac{1}{2J_i+1}\left|\sum_{L}\braket{f||\Xi_{JL}\left(\kappa_e,\kappa_\nu\right)||i}\right|^2,
\end{align}
where $J_i$ is the angular momentum of the initial nuclear state
and $E_0$ is the maximum energy of the electron.
The nonzero integers $\kappa_e$ and $\kappa_\nu$ 
 represent the lepton angular momentum for the electron and neutrino, respectively, with 
\begin{align}
    j_\kappa=|\kappa|-1/2, \hspace{1cm}
    l_\kappa=
    \begin{cases}
    \kappa & (\kappa>0) \\
    -\kappa-1 & (\kappa<0)
    \end{cases}.
\end{align}
The bracket $\braket{f||\Xi_{JL}||i}$ denotes the reduced matrix element (RDME) \cite{sakurai1985modern} of the multipole operator $\Xi_{JL}$,
which will be detailed in the next subsection.

\subsection{Multipole operator}

As shown in Eq.~(12) of Ref.~\cite{horiuchi2021electron}, the multipole operator $\Xi_{JL}$ is written down as
\begin{align}
  \Xi_{JLM}(\kappa_e,\kappa_\nu) & =  S_{\kappa_e}\int d\bm{r} \nonumber \\
  &\times  \left\{
  \mp Y_{JM}(\hat{r})V_0(\bm{r})\delta_{L,J}
                    \Phi_{0JJ}(\kappa_e,\kappa_\nu;r)\right.
             \pm i [Y_L(\hat{r})\otimes\bm{V}(\bm{r})]_{JM}
                    \phi_{1LJ}(\kappa_e,\kappa_\nu;r)
        \notag \\
        &   +i Y_{JM}(\hat{r})A_0(\bm{r})\delta_{L,J}
                    \phi_{0JJ}(\kappa_e,\kappa_\nu;r)
           - \left.[Y_L(\hat{r})\otimes \bm{A}(\bm{r})]_{JM}
                    \Phi_{1LJ}(\kappa_e,\kappa_\nu;r)
      \right\}, \label{eq:xi}
\end{align}
where the plus-minus signs correspond to the $\beta^\mp$ decay.
Here, $S_\kappa=\textrm{sgn}(\kappa)$ is the sign of $\kappa$ and
\begin{align}
  \Phi_{KLJ}(\kappa_e,\kappa_\nu;r) & = 
        G_{\kappa_e}(r)g_{\kappa_\nu}(r)S_{KLJ}(\kappa_e,\kappa_\nu)
      + F_{\kappa_e}(r)f_{\kappa_\nu}(r)S_{KLJ}(-\kappa_e,-\kappa_\nu),\\
  \phi_{KLJ}(\kappa_e,\kappa_\nu;r) & =  
        G_{\kappa_e}(r)f_{\kappa_\nu}(r)S_{KLJ}(\kappa_e,-\kappa_\nu)
      - F_{\kappa_e}(r)g_{\kappa_\nu}(r)S_{KLJ}(-\kappa_e,\kappa_\nu)
\end{align}
with
\begin{align}
    S_{KLJ}\left(\kappa',\kappa\right)=&\sqrt{2\left(2j_\kappa+1\right)\left(2j_{\kappa'}+1\right)\left(2l_\kappa+1\right)\left(2l_{\kappa'}+1\right)(2K+1)} \nonumber\\
    &\times \left(l_\kappa,0,l_{\kappa'},0|L,0\right)
    \begin{Bmatrix}
        l_{\kappa'} & 1/2 & j_{\kappa'} \\
        l_{\kappa} & 1/2 & j_{\kappa} \\
        L & K & J
    \end{Bmatrix}.
\end{align}
Here $(j_1,m_1,j_2,m_2|J,M)$ is the Clebsch--Gordan coefficient
and the curly brackets denote the 9-j symbol~\cite{rose1957ang, edmonds1996angular}. 
The radial wave functions of the emitted electron or the positron,
$G_\kappa(r)$ and $F_\kappa(r)$, are obtained by solving the radial Dirac equation with the Coulomb potential $V_C(r)$ as
\begin{align}
    \frac{dG_\kappa(r)}{dr}+\frac{1+\kappa}{r}G_\kappa(r)-\left(m_e+E_e-V_C(r)\right)F_\kappa(r)=0, \label{eq:Dirac1}\\
    \frac{dF_\kappa(r)}{dr}+\frac{1-\kappa}{r}F_\kappa(r)-\left(m_e-E_e+V_C(r)\right)G_\kappa(r)=0. \label{eq:Dirac2}
\end{align}
On the other hand, the neutrino is treated as a plane wave with the momentum
$p_\nu = E_0 - E_e$,
and its radial wave functions are given by $g_\kappa(r)=j_{l_\kappa}\left(p_\nu r\right)$ and $f_\kappa(r)=S_\kappa j_{l_{-\kappa}}\left(p_\nu r\right)$, where
$j_l(x)$ is the $l$th-order spherical Bessel function.

For a case of the angular momentum $J$ with the parity $P = (-1)^{J+1}$
such as $J^P=0^-$, $1^+$, and $2^-$, the multipole operator is written as
\begin{align}
  \sum_L \Xi_{JLM}(\kappa_e,\kappa_\nu)  &=  S_{\kappa_e}\int d\bm{r}
   \Big\{
  \pm i [Y_J(\hat{r})\otimes\bm{V}(\bm{r})]_{JM} \phi_{1JJ}(\kappa_e,\kappa_\nu;r)
     \notag\\
 & +   i Y_{JM}(\hat{r})A_0(\bm{r})\phi_{0JJ}(\kappa_e,\kappa_\nu;r)\notag\\
 &    - \sum_{L = J\pm 1}[Y_L(\hat{r})\otimes \bm{A}(\bm{r})]_{JM}
                    \Phi_{1LJ}(\kappa_e,\kappa_\nu;r)
   \Big\}.  \label{eq:xi-un}
\end{align}
The spatial component of the vector current (1st term) is 
the `magnetic' operator, while that of the axial-vector current (3rd term) 
is the `electric' operator. 
The time component  (2nd term)
and the spatial component (3rd term) of the axial vector current
contribute to the $0^-$ operator.

For a case of the angular momentum $J$ with parity $P = (-1)^{J}$
such as $J^P=0^+$, $1^-$, and $2^+$, the multipole operator is written as
\begin{align}
 \sum_L \Xi_{JLM}(\kappa_e,\kappa_\nu) & =  S_{\kappa_e}\int d\bm{r}\Big\{
  \mp Y_{JM}(\hat{r})V_0(\bm{r})
                    \Phi_{0JJ}(\kappa_e,\kappa_\nu;r)\notag\\
                    &\pm i \sum_{L = J \pm 1}[Y_L(\hat{r})\otimes\bm{V}(\bm{r})]_{JM}
                    \phi_{1LJ}(\kappa_e,\kappa_\nu;r)
        \notag \\
        &              - [Y_J(\hat{r})\otimes \bm{A}(\bm{r})]_{JM}
                    \Phi_{1JJ}(\kappa_e,\kappa_\nu;r)
      \Big\}. \label{eq:xi-n}
\end{align}
In this case, the spatial component of the vector current (2nd term) is the `electric' operator, 
while the spatial component of the axial-vector current (3rd term) 
is the `magnetic' operator.
For the $0^+$ operator, the time component (1st term)
and the spatial component (2nd term) of the vector current 
contribute to the transition.

\subsection{Longitudinal component of the vector current}

The  meson-exchange current (MEC)
for the spatial component of the isovector vector current
plays an important role, while the impulse approximation works well for the time component of the vector current in most cases.
Since it is rather hard to construct many-body nuclear currents that satisfy the current conservation relation consistent
with the nuclear wave function obtained from the effective nuclear potential,
it is effective to rewrite a part of the spatial
component of the vector current in terms of the time component
of the vector current and use the impulse approximation.
Here we focus on the second term of Eq.~(\ref{eq:xi-n}) for the `electric' transition.
The first term of Eq.~(\ref{eq:xi-un}) has nothing to do with the current conservation law.

For the nuclear beta decay, due to the Coulomb distortion of the electron wave function, 
it is not straightforward to implement the current conservation relation
in contrast to the plane-wave formula for neutrino reaction and beta decay rate~\cite{walecka2012semileptonic,Nakamura:2000vp}.
Here we use the Bessel--Fourier transformation to take advantage of the momentum space representation.  
The relevant terms of the multipole operator are written as

\begin{align}
  \sum_{L = J \pm 1}[Y_L(\hat{r})\otimes\bm{V}(\bm{r})]_{JM}
  \phi_{1LJ}(\kappa_e,\kappa_\nu;r) & = 
\frac{2}{\pi}\int_0^\infty dq\,q^2  \ {\cal O}^J(\kappa_e,\kappa_\nu;q),
   \label{eq:vlon1}
\end{align}
where
\begin{align}
{\cal O}^J(\kappa_e,\kappa_\nu;q)  & =   
T^{JJ-1M}(q) \tilde{\phi}_{1J-1J}(\kappa_e,\kappa_\nu;q) + T^{JJ+1M}(q) \tilde{\phi}_{1J+1J}(\kappa_e,\kappa_\nu;q).
\label{eq:oorig}
\end{align}
Here we used the identity,
\begin{eqnarray}
 \frac{2}{\pi}\int_0^\infty dq q^2 j_L(qr)j_L(qr') & = & \frac{\delta(r-r')}{r^2}.
 \label{eq:orthogonality_of_sphericalBessel}
\end{eqnarray}
The momentum-space multipole operator and lepton wave function
are respectively defined as
\begin{align}
  T^{JLM}(q) & =  \int d\bm{r}\,
  j_{L}(qr)[Y_{L}(\hat{r})\otimes \bm{V}(\bm{r}) ]_{JM}, \\
  \tilde{\phi}_{1LJ}(\kappa_e,\kappa_\nu;q) & = 
  \int_0^\infty dr\,r^2 j_L(qr) \phi_{1LJ}(\kappa_e,\kappa_\nu;r) .
\end{align}
Hereafter, $\tilde{\phi}_{1LJ}(\kappa_e,\kappa_\nu;q) $ is simplified as
$\tilde{\phi}_L(q)$.

 The electric $T_E^{JM}$
and longitudinal $T_L^{JM}$ multipole operators are defined by
\begin{align}
  T_E^{JM}(q) & =  \frac{1}{q} \int d\bm{r}\,
  \bm{\nabla}\times [j_J(qr) \bm{Y}_{JJM}(\hat{r})]\cdot \bm{V}(\bm{r})
  \nonumber \\
  & = 
  -i \sqrt{\frac{J}{2J+1}} T^{JJ+1M}(q)
 + i \sqrt{\frac{J+1}{2J+1}}  T^{JJ-1M}(q),
 \label{eq:elec}\\
  T_L^{JM}(q) & =  \frac{i}{q} \int d\bm{r}\,
  \bm{\nabla} [j_J(qr) Y_{JM}(\hat{r})]\cdot \bm{V}(\bm{r})
  \nonumber \\
& =    i \sqrt{\frac{J+1}{2J+1}} T^{JJ+1M}(q)
  + i \sqrt{\frac{J}{2J+1}} T^{JJ-1M}(q).
\label{eq:long}
\end{align}
Note that in Ref.~\cite{walecka2012semileptonic} $T_E^{JM}$ and $T_L^{JM}$ 
are called $T_{JM}^{el}$ and $L_{JM}$, respectively.
Using these longitudinal and electric multipole operators, Eq.~(\ref{eq:oorig}) can be written as
\begin{align}
  i{\cal O}^J(\kappa_e,\kappa_\nu;q) & = 
    T_E^{JM}(q)\left(\sqrt{\frac{J+1}{2J+1}}\tilde{\phi}_{J-1}(q)
    -\sqrt{\frac{J}{2J+1}}\tilde{\phi}_{J+1}(q)\right)
    \nonumber \\
    & + 
    T_L^{JM}(q)\left(\sqrt{\frac{J}{2J+1}}\tilde{\phi}_{J-1}(q)
    + \sqrt{\frac{J+1}{2J+1}}\tilde{\phi}_{J+1}(q)\right).
\end{align}

We incorporate the current conservation relation in Eq. (\ref{eq:oorig}).
The current conservation relation of the vector current of the
charged current is expressed by
\begin{align}
  \bm{\nabla}\cdot\bm{V}+i[H,V_0]=0, \label{eq:ccx}
\end{align}
where the nuclear Hamiltonian $H$ is assumed to be  isospin invariant.
To include the Coulomb interaction for protons and mass difference between
proton and neutron, the extended Siegert's theorem has to be used.
In the momentum space, the relation is expressed in terms of
the longitudinal multipole operator $T_L^{JM}(q)$
and the Coulomb multipole operator $T_C^{JM}(q)$ as
\begin{align}
 q T_L^{JM}(q) + [H , T_C^{JM}(q)]  =  0. \label{eq:loco}
\end{align}
Here,
\begin{align}
  T_C^{JM}(q) =  \int d\bm{r}\,j_J(qr) Y_{JM}(\hat{r}) V_0(\bm{r}),
\end{align}
which is called $M_{JM}$ in Ref.~\cite{walecka2012semileptonic}.
When we take the nuclear matrix element of the operator,
the commutator with the Hamiltonian is reduced to
the energy difference between the
final and initial nuclear states $\omega = E_f - E_i $,
which corresponds to the
maximum energy of electron $E_0 = -\omega$ .
Then we obtain
\begin{align}
  T_L^{JM}(q) & =  - \frac{\omega}{q}T_C^{JM}(q).
\end{align}
Furthermore, the term $T^{JJ-1M}(q)$, which is 
the leading term of the electric multipole operator $T_E^{JM}(q)$
in the long-wave-length limit, can be written as 
\begin{align}
  T^{JJ-1M}(q) & =  i\sqrt{\frac{2J+1}{J}}\frac{\omega}{q}T_C^{JM}(q)-\sqrt{\frac{J+1}{J}}T^{JJ+1M}(q). \label{eq:lo-co}
\end{align}
Inserting Eq.~\eqref{eq:lo-co} to Eq.~\eqref{eq:oorig}, we get
\begin{align}
  i{\cal O}^J(\kappa_e,\kappa_\nu;q)  & = 
    - \sqrt{\frac{2J+1}{J}}\frac{\omega}{q}T_C^{JM}(q)\tilde{\phi}_{J-1}(q)\notag\\
 &+ i T^{JJ+1M}(q)\left(\tilde{\phi}_{J+1}(q)
  - \sqrt{\frac{J+1}{J}}\tilde{\phi}_{J-1}(q)\right),
\end{align}
for $J\ge 1$.

For $J=0$, ${\cal O}^J$ can be expressed by the
Coulomb form factor as
\begin{eqnarray}
  i{\cal O}^0(\kappa_e,\kappa_\nu;q)  & = &
    - \frac{\omega}{q}T_C^{00}(q)\tilde{\phi}_{1}(q).
\end{eqnarray}

Finally, we transform the momentum space expression back to the coordinate
space, using Eq.~\eqref{eq:orthogonality_of_sphericalBessel} and the following well-known identities:
\begin{align}
  \int_0^\infty dq\,q j_J(qr) j_{J-1}(qr') & = \frac{\pi}{2}\frac{{r'}^{J-1}}{r^{J+1}}\theta(r-r'),\\
  \int_0^\infty dq\,q^2 j_{J+1}(qr) j_{J-1}(qr')& =  \frac{\pi}{2}\left[
    (2J+1)\frac{{r'}^{J-1}}{r^{J+2}}\theta(r-r')- \frac{\delta(r-r')}{r^2}\right].
\end{align}
The multipole operator for $P = (-1)^{J}$ ($J \ge 1$) 
is given as
\begin{align}
&\sum_L \Xi_{JLM}(\kappa_e,\kappa_\nu)  =  S_{\kappa_e}\int d\bm{r}\,
   \left\{
  \mp Y_{JM}(\hat{r})V_0(\bm{r})
  \left(\Phi_{0JJ}(\kappa_e,\kappa_\nu;r)+
  \omega \sqrt{\frac{2J+1}{J}}\frac{\varphi_J(r)}{r^{J+1}}\right)\right.\nonumber \\
&\pm i [Y_{J+1}(\hat{r})\otimes\bm{V}(\bm{r})]_{JM}
  \left[\phi_{1J+1J}(\kappa_e,\kappa_\nu;r) 
 +\sqrt{\frac{J+1}{J}}\left(\phi_{1J-1J}(\kappa_e,\kappa_\nu;r)
-(2J+1)\frac{\varphi_J(r)}{r^{J+2}}\right)\right]
        \notag \\
&- [Y_J(\hat{r})\otimes \bm{A}(\bm{r})]_{JM}
                    \Phi_{1JJ}(\kappa_e,\kappa_\nu;r)
      \Bigg\}, \label{eq:xi-n-v2}
\end{align}
where
\begin{align}
\varphi_J(r)  =  \int_0^r dr'\,(r')^{J+1}\phi_{1J-1J}(\kappa_e,\kappa_\nu;r').
\end{align}
The magnetization current,
$\bm{V} \sim \bm{\nabla} \times \bm{\mu}$,
is purely transverse, which gives
$T_L^{JM}(q)=0$ and $T^{JJ-1M}(q) = - \sqrt{(J+1)/J}T^{JJ+1M}(q)$.
The relation shows that the 
expression for the magnetization current in  Eq.~(\ref{eq:xi-n})
is equivalent to that of Eq.~(\ref{eq:xi-n-v2}). 

For $J=0$, 
the multipole operator is written as
\begin{eqnarray}
  \Xi_{000}(\kappa_e,\kappa_\nu) & = &
  \mp S_{\kappa_e}\int d\bm{r}  Y_{00}(\hat{r})V_0(\bm{r})\varphi_0(r)
\end{eqnarray}
with
\begin{eqnarray}
\varphi_0(r) & = &  \Phi_{000}(\kappa_e,\kappa_\nu;r) +  \omega \int_r^\infty dr'
  \phi_{110}(\kappa_e,\kappa_\nu;r'). \label{eq:coul0}
\end{eqnarray}
The second term of Eq. (\ref{eq:coul0}) is the contribution
of the longitudinal term of the spatial component of the vector current.

For the transition to the isobaric analog state (IAS),
$\omega=0$  assuming  the isospin invariance of the nuclear Hamiltonian
and we obtain
\begin{eqnarray}
  \varphi_0(r) = \Phi_{000}(\kappa_e,\kappa_\nu;r).
\end{eqnarray}
The formula shows that the decay rate  is 
given by the time component of the vector current alone.  
There is no need to
include the correction originating from the spatial component of the vector current.

The $0^+$ transition to the non-IAS states is weak compared to the
transition to the IAS because of the orthogonality relation
$\braket{f|\int d\bm{r} V^0|i} = \braket{f|T^{\pm}|i}=0$.
For practical computations, it is convenient to rewrite Eq. (\ref{eq:coul0}) using this orthogonality relation.
For the transition to the non-IAS, we can use $\varphi_0(r)$ subtracted by a
constant value as
\begin{eqnarray}
  \varphi_0(r) & \rightarrow &  \varphi_0(r) - \varphi_0(0).
\end{eqnarray}
Thus, we obtain
\begin{eqnarray}
  \varphi_0(r) = \Phi_{000}(\kappa_e,\kappa_\nu;r) - \Phi_{000}(\kappa_e,\kappa_\nu;0)
         - \omega \int_0^r dr'
  \phi_{110}(\kappa_e,\kappa_\nu;r').\label{eq:coul1}
\end{eqnarray}
An advantage of Eq. (\ref{eq:coul1}) is that $\varphi_0(r)$ 
is an operator of the form of `second-forbidden transition'
starting from ${\cal O}(r^2)$
and the integration of $r$ to $r \rightarrow \infty$ of the $\omega$ term  is avoided.

\section{Impulse approximation}
\label{sec:impilse_approx}

In this section, we present  explicit expressions of the multipole operator
in the impulse approximation of the nuclear current.
For the axial-vector current ${\cal A}$, the space and time components of 
the one-body charged-current operators
in non-relativistic approximation are respectively given as
\begin{align}
  \bm{A}(\bm{r})& = \sum_{j=1}^A  \delta(\bm{r} - \bm{r}_j)\left[
    g_A(q^2) \bm{\sigma}_j 
    - g_P(q^2) \frac{\bm{q} (\bm{\sigma}_j\cdot\bm{q})}{2M}
    \right]\tau^{\pm}_j,   \\
   A_0(\bm{r})& = \sum_{j=1}^A \delta(\bm{r} - \bm{r}_j)\left[
   g_A(q^2) \frac{\bm{\sigma}_j \cdot (\bm{p}_j' + \bm{p}_j) }{2M}
   - g_P(q^2)\frac{\omega (\bm{\sigma}_j\cdot\bm{q})}{2M}\right]\tau^{\pm}_j,
\end{align}
where $\bm{p}_j'$ and $\bm{p}_j$ are the momentum operators
of the $j$th nucleon for the final and initial states, respectively,
and $\bm{q}$ is the momentum transfer from the lepton,
$\bm{q} = \bm{p}_j' - \bm{p}_j$.
Here the nucleon form factors $g_A(q^2)$ and $g_P(q^2)$ are
functions of momentum transfer $q^2$ to the nucleon. 
In the nuclear beta decay, the use of the form factors at $q^2 = 0$ is a good approximation: $g_A(0)=1.27$ and $g_P(0) = g_A(0) 2M/m_\pi^2 $
assuming the pion-pole dominance~\cite{fearing2003}.

For the vector current ${\cal V}$, the operators are given as
\begin{align}
  \bm{V}(\bm{r})& = \sum_{j=1}^A \delta(\bm{r} - \bm{r}_j)\left[
   g_V(q^2) \frac{ (\bm{p}_j' + \bm{p}_j) }{2M}
+ \frac{g_V(q^2) + g_M(q^2)}{2M} i \bm{\sigma}_j \times \bm{q} \right]\tau^{\pm}_j, \\
   V_0(\bm{r})& = \sum_{j=1}^A  \delta(\bm{r} - \bm{r}_j)g_V(q^2)\tau^{\pm}_j,
\end{align}
where $g_V(0)=1$, $g_V(0)+g_M(0) = \mu_p - \mu_n = 4.7$.
Hereafter,  we simply write 
the form factors at $q^2=0$ as $g_A, g_V, g_M$ and $g_P$.

  The $g_A\bm{\sigma}\tau^\pm$ term of the spatial component
  of the axial-vector current and
$g_V \tau^\pm$ term of 
  the time component of the vector current are main contributors
  remaining in the non-relativistic limit of the nucleon current.
 The $g_A\bm{\sigma}\cdot(\bm{p}+\bm{p}')/2M$
 term of the time component of the axial-vector current
 and   the $g_V (\bm{p}+\bm{p}')/2M$ term of 
 the spatial component of the vector current (convection current),
 are  velocity-dependent $(v/c)$ correction terms.
The $(g_V+g_M)/(2M)$ term is the induced weak-magnetism term.

\subsection{Case of $P = (-1)^{J+1}$}

In the impulse approximation, the multipole operator for $P = (-1)^{J+1}$ is given as
\begin{align}
  \sum_L \Xi_{JLM}(\kappa_e,\kappa_\nu) & =  S_{\kappa_e}\int d\bm{r}\notag\\
  &\times \Bigg[\sum_{j=1}^A \tau_j^\pm \delta(\bm{r} - \bm{r}_j)\Big\{
              [Y_{J-1}(\hat{r})\otimes \bm{\sigma}_j]_{JM}\phi_a(r)
    + [Y_{J+1}(\hat{r})\otimes \bm{\sigma}_j]_{JM}\phi_b(r)
    \nonumber \\
    &  
    + Y_{JM}\bm{\sigma}_j\cdot\bm{\nabla}_j \phi_c(r)
    + [Y_{J}(\hat{r})\otimes \bm{\nabla}_j]_{JM}\phi_d(r)\Big\}\Bigg],
    \label{eq:ia-unp}
\end{align}
with
\begin{align}
  \phi_a(r) & = 
  - g_A \Phi_{1 J-1 J}(r)
 +  \frac{1}{2M}\sqrt{\frac{J}{2J+1}}D_-^J 
 \left[g_A \phi_{0JJ}(r) + g_P m_e \bar{\phi}_{0JJ}(r)\right]\notag\\
& \mp \frac{g_V + g_M}{2M}\sqrt{\frac{J+1}{2J+1}}D_-^J \phi_{1JJ}(r), \label{eq:phi_a}\\
  \phi_b(r) & = 
  - g_A \Phi_{1 J+1 J}(r)
 -  \frac{1}{2M}\sqrt{\frac{J+1}{2J+1}}D_+^J \left[g_A \phi_{0JJ}(r)
  + g_P m_e \bar{\phi}_{0JJ}(r)\right]\notag\\
 &\mp \frac{g_V + g_M}{2M}\sqrt{\frac{J}{2J+1}}D_+^J \phi_{1JJ}(r), \label{eq:phi_b}\\
 \phi_c(r) & =  \frac{g_A}{M}\phi_{0JJ}, \label{eq:phi_c}\\
 \phi_d(r) & =  \pm \frac{g_V}{M}\phi_{1JJ}(r), \label{eq:phi_d}
\end{align} 
where
\begin{align}
  D_+^J & =  \frac{d}{dr} - \frac{J}{r}, \\
  D_-^J & =  \frac{d}{dr} + \frac{J+1}{r},\\
  \bar{\phi}_{0JJ}(r) & = 
        G_{\kappa_e}(r)f_{\kappa_\nu}(r)S_{0JJ}(\kappa_e,-\kappa_\nu)
      + F_{\kappa_e}(r)g_{\kappa_\nu}(r)S_{0JJ}(-\kappa_e,\kappa_\nu).
\end{align}
The simplified notations $\Phi_{KLS}(r)$ and $\phi_{KLS}(r)$ denoting
$\Phi_{KLS}(\kappa_e,\kappa_\nu;r)$ and $\phi_{KLS}(\kappa_e,\kappa_\nu;r)$ respectively are used.

The first terms of $\phi_a$ (Eq.~(\ref{eq:phi_a}))
  and $\phi_b$ (Eq.~(\ref{eq:phi_b})) 
  are contributions of the spatial component of the axial-vector current,
  while the $g_A/M$ terms of  $\phi_a$, $\phi_b$ and $\phi_c$ (Eq. (\ref{eq:phi_c}))
 are contributions of the time component of the axial vector current.
The last terms, $(g_V+g_M)/(2M)$, of $\phi_a$ and $\phi_b$
are the weak-magnetism terms. The $\phi_d$ (Eq. (\ref{eq:phi_d}))
is the contribution of the convection current.

 The Dirac equation, Eqs.~\eqref{eq:Dirac1} and \eqref{eq:Dirac2}, 
 is used to derive the $g_P$ term in Eqs.~\eqref{eq:phi_a} and \eqref{eq:phi_b}. 
 Here only $m_e$ remains and the Coulomb potential $V_C(r)$ disappears due to the
 gauge invariance~\cite{BLOKHINTSEV1962498}.
The $g_P$ term
is usually neglected for the nuclear beta decay~\cite{sch66} because
the contributions of the induced pseudo-scalar term $g_P$ 
are about ${\cal O}(m_e E_0/m_\pi^2)$, typically $\sim 10^{-3}$, smaller than those of the leading $g_A$ terms of $\phi_a$ and $\phi_b$. 

\subsection{Case of $P = (-1)^{J}$}

In the impulse approximation, the multipole operator for $P = (-1)^{J}$ is given as
\begin{align}
  \sum_L \Xi_{JLM}(\kappa_e,\kappa_\nu) & = S_{\kappa_e}\int d\bm{r}\notag\\
  &\times\Bigg[\sum_{j=1}^A \tau_j^\pm \delta(\bm{r} - \bm{r}_j) \Big\{
              Y_{JM}(\hat{r})\phi_A(r)
    + [Y_{J}(\hat{r})\otimes \bm{\sigma}_j]_{JM}\phi_B(r)
    \nonumber \\
    &
    +  [Y_{J+1}(\hat{r})\otimes \bm{\nabla}_j]_{JM}\phi_C(r)\Big\}\Bigg]
\end{align}
with
\begin{align}
  \phi_A(r) & = 
  \mp g_V\left(\Phi_{0 J J}(r) +
  \omega \sqrt{\frac{2J+1}{J}}\frac{\varphi_J(r)}{r^{J+1}} -\frac{1}{2M}\sqrt{\frac{J+1}{J}}\psi_J(r)\right), \label{eq:phi_la}\\ 
  \phi_B(r) & = 
  - g_A \Phi_{1 J J}(r)
  \mp \frac{g_V + g_M}{2M}\psi_J(r), \label{eq:phi_lb}\\
  \phi_C(r) & =  \pm \frac{g_V}{M}\left\{\phi_{1J+1J}(r)
   +\sqrt{\frac{J+1}{J}}\left[\phi_{1J-1J}(r)
   - (2J+1)\frac{\varphi_J(r)}{r^{J+2}}\right]\right\},\label{eq:phi_lc}
\end{align} 
and
\begin{align}
   \psi_J(r)&=\sqrt{\frac{J+1}{2J+1}}D_+^{J-1} \phi_{1J-1J}(r)
  + \sqrt{\frac{J}{2J+1}}D_-^{J+1} \phi_{1J+1J}(r).
\end{align} 
The first terms of $\phi_A$ (Eq. (\ref{eq:phi_la})) and $\phi_B$
  (Eq. (\ref{eq:phi_lb}))
  are the contribution of the time component of the vector current
  and the space component of the axial-vector current, respectively.
  The second term of $\phi_B$ is the weak-magnetism term.
  The second and third terms of  $\phi_A$ and
  the $\phi_C$ (Eq. (\ref{eq:phi_lc})) are contributions of the convection
  current. The second term of $\phi_A$,
  which is proportional to $\omega$, is usually taken into account
  in the leading order approximation of the $\varphi_J$.

\section{Explicit formulas and applications}
\label{sec:applications}

In this section, we give explicit formulas for the allowed $J^P=1^+$ and 
the first-forbidden $J^P=0^-$, $1^-$, and $2^-$ transitions 
useful in practical nuclear-structure calculations.
We include all the recoil order currents ${\cal O}(p/M)$ that are considered as higher-order corrections to the Fermi and Gamow--Teller operators.
We show that these formulas are reduced into the
  widely-used leading-order formula by
  Behrens--B\"{u}hring (LOB)\cite{Behrens:1971rbq}
with the approximation of the electron and neutrino wave functions
described in Eqs. (60)--(65) of paper (I).
We then apply our formulas to the beta-decay rate of $^{160}$Sn, 
where the first-forbidden transitions, in particular the $J^P=1^-$ transition, 
are expected to give a sizable contribution to the total beta-decay
rate~\cite{mus16}.

The RDME of the multipole operator is 
given by the radial integral of the transition densities 
together with the lepton wave functions.
For $P=(-1)^{J+1}$, the RDME is given as

\begin{align}
  \braket{f|| \sum_L \Xi_{JL}(\kappa_e,\kappa_\nu)||i}
  & = S_{\kappa_e}
  \int_0^\infty  dr\, r^2 \notag\\
  &\times\left[\rho_{J-1J}^\sigma(r)\phi_a(r)
              + \rho_{J+1J}^\sigma(r)\phi_b(r)
              + \rho_J^{\sigma\nabla}(r)\phi_c(r)
              + \rho_{JJ}^\nabla(r)\phi_d(r) \right], 
\end{align}
and for $P=(-1)^{J}$
\begin{align}
  \braket{f|| \sum_L \Xi_{JL}(\kappa_e,\kappa_\nu)||i}
  & =  S_{\kappa_e}
  \int_0^\infty  dr\, r^2
  \left[\rho_{J}(r)\phi_A(r)
              + \rho_{JJ}^\sigma(r)\phi_B(r)
              + \rho_{J+1J}^\nabla(r)\phi_C(r) \right],
\end{align}
where the transition densities used in the impulse approximation
are defined as
\begin{align}
  \rho_{J}(r)
  & =  \braket{f||\sum_{j=1}^A  \int d\Omega_r \delta(\bm{r}-\bm{r}_j)\tau^{\pm}_j
                 {}Y_J(\hat{r}) ||i},\\
  \rho_{LJ}^\sigma(r)
  & =  \braket{f||\sum_{j=1}^A \int d\Omega_r \delta(\bm{r}-\bm{r}_j)\tau^{\pm}_j
                 {}[Y_L(\hat{r})\otimes\bm{\sigma}_j]_J ||i},\\
  \rho_{LJ}^\nabla(r)
  & =  \braket{f||\sum_{j=1}^A \int d\Omega_r \delta(\bm{r}-\bm{r}_j)\tau^{\pm}_j
                 {}[Y_L(\hat{r})\otimes\bm{\nabla}_j]_J ||i},\\
  \rho_{J}^{\sigma\nabla}(r)
  & =  \braket{f||\sum_{j=1}^A \int d\Omega_r \delta(\bm{r}-\bm{r}_j)\tau^{\pm}_j
                 {}Y_J(\hat{r}) \bm{\sigma}_j\cdot\bm{\nabla}_j ||i}.
\end{align}  
Here, $\int d\Omega_r$ denotes the integration over solid angles.
We employ a nuclear EDF method to generate these transition 
densities. As the details can be found in Ref.~\cite{yos13}, 
here we recapitulate the numerical procedures relevant 
to the present analysis.
In the framework of the nuclear EDF method,
the ground state of a mother nucleus is described by solving the Kohn--Sham--Bogoliubov (KSB) equation~\cite{dob84}.
We solve the KSB equation in cylindrical coordinates 
assuming the axial symmetry~\cite{kas21}.
Then, the transitions to a daughter nucleus are
described by the proton--neutron quasiparticle-random-phase approximation (pnQRPA).
The residual interactions entering into the pnQRPA equation
are given by the EDF self-consistently. 
The EDF consists of the Skyrme SLy4 functional~\cite{cha98} and 
the Yamagami--Shimizu--Nakatsukasa functional~\cite{yam09} for the particle-hole 
and pairing energy, respectively.
Since we assume the axial symmetry, the $z$-component of 
the total angular momentum $J_z=K$ is a good quantum number.

In the actual computations, we evaluate the following 
transition densities:
\begin{align}
    \delta \rho^{1}(\br)&
    :=\braket{ f|\sum_{\sigma, \sigma^\prime, \tau, \tau^\prime}\psi^\dagger(\br\sigma^\prime \tau^\prime)\langle \sigma^\prime|1|\sigma\rangle \langle\tau^\prime|\tau^\pm|\tau\rangle\psi(\br \sigma \tau)|i},\\
    \delta \rho^{\sigma}_{\mu_\sigma}(\br)&
    :=\braket{ f|\sum_{\sigma, \sigma^\prime, \tau, \tau^\prime}\psi^\dagger(\br\sigma^\prime \tau^\prime)\langle\sigma^\prime|\sigma_{\mu_\sigma}|\sigma\rangle \langle\tau^\prime|\tau^\pm|\tau\rangle\psi(\br \sigma \tau)|i},\\
    \delta \rho^{\nabla}_{\mu_r}(\br)&  
    :=\braket{ f|\sum_{\sigma, \sigma^\prime, \tau, \tau^\prime}\psi^\dagger(\br\sigma^\prime \tau^\prime)\langle\sigma^\prime|1|\sigma\rangle \langle\tau^\prime|\tau^\pm|\tau\rangle\nabla_{\mu_r}\psi(\br \sigma \tau)|i},\\
    \delta \rho^{\sigma \nabla}(\br)&
    := \bra{ f}|
      \sum_{\sigma, \sigma^\prime, \tau, \tau^\prime}
     \sum_{\mu_\sigma,\mu_r} 
      \psi^\dagger(\br\sigma^\prime \tau^\prime)\langle\sigma^\prime|\sigma_{\mu_\sigma}|\sigma\rangle\langle\tau^\prime|\tau^\pm|\tau\rangle\nabla_{\mu_r}\psi(\br \sigma \tau) \notag \\
    &\times (1,\mu_\sigma,1,\mu_r|0,0)\ket{ i} ,
\end{align}
where $\sigma_{\mu_\sigma}$ and $\nabla_{\mu_r}$ are spherical components of 
the Pauli matrix $\boldsymbol{\sigma}$ and the operator $\boldsymbol{\nabla}$, 
respectively. With these, we can evaluate the transition 
densities appearing in the RDMEs as
\begin{align}
    \rho_{JK}(r)
    &=\int d\Omega_r \delta \rho^1(\br)Y_{JK}(\hat{r}),\\
    \rho_{LJK}^\sigma(r)
    &=
    \int d\Omega_r
    \sum_{\mu_\sigma,M}
    \delta \rho^\sigma_{\mu_\sigma}(\br)Y_{LM}(\hat{r})(L,M,1,\mu_\sigma|J,K),\\
    \rho_{LJK}^\nabla(r)
    &=
    \int d\Omega_r
   \sum_{\mu_r,M}
    \delta \rho^\nabla_{\mu_r}(\br)Y_{LM}(\hat{r})(L,M,1,\mu_r|J,K),\\
    \rho_{JK}^{\sigma \nabla}(r)
    &=\int d\Omega_r \delta \rho^{\sigma \nabla}(\br)Y_{JK}(\hat{r}).
\end{align}

For spherical nuclei, we only need to evaluate the 
transition densities for $K=0$. 
Then, one obtains the transition densities  as
\begin{align}
    \rho_J(r) &= \sqrt{2J+1}\rho_{J0}(r),\\
    \rho_{LJ}^\sigma(r) &= \sqrt{2J+1}\rho^\sigma_{LJ0}(r),\\
    \rho_{LJ}^\nabla(r) &= \sqrt{2J+1}\rho^\nabla_{LJ0}(r),\\
    \rho_{J}^{\sigma \nabla}(r) &= \sqrt{2J+1}\rho^{\sigma \nabla}_{J0}(r)
\end{align}
for any $L$ and $J$.

\subsection{Case for $J^P=1^+$}

An explicit formula of the multipole operator for the allowed
Gamow--Teller transition in the impulse approximation is given as
\begin{align}
  \braket{f|| \sum_L \Xi_{1L}(\kappa_e,\kappa_\nu)||i}
  & = 
S_{\kappa_e}   \int_0^\infty dr\, r^2 \notag\\
&\times\left[
               \rho_{01}^\sigma(r)\phi_a(r)
              +  \rho_{21}^\sigma(r)\phi_b(r)
               + \rho_1^{\sigma\nabla}(r)\phi_c(r)
               + \rho_{11}^{\nabla}(r)\phi_d(r)
                \right],
\end{align}
where 
\begin{align}
  \phi_a(r) & =  -g_A \Phi_{101}(r)
  + \frac{1}{2M}\sqrt{\frac{1}{3}}\left(\frac{d}{dr}+\frac{2}{r}\right)
  \left[g_A\phi_{011}(r) + g_P m_e\bar{\phi}_{011}(r)\right]\notag\\
  &\mp \frac{g_V+g_M}{2M}\sqrt{\frac{2}{3}}\left(\frac{d}{dr}+\frac{2}{r}\right)\phi_{111}(r)
  , \label{eq:gtphia}\\
  \phi_b(r) & =  -g_A \Phi_{121}(r)
  - \frac{1}{2M}\sqrt{\frac{2}{3}}\left(\frac{d}{dr}-\frac{1}{r}\right)
 {} [g_A \phi_{011}(r)+ g_P m_e\bar{\phi}_{011}(r)] \notag\\
  &\mp \frac{g_V+g_M}{2M}\sqrt{\frac{1}{3}}\left(\frac{d}{dr}-\frac{1}{r}\right)\phi_{111}(r)
  , \label{eq:gtphib}\\
 \phi_c(r) & =  \frac{g_A}{M}\phi_{011}(r), \label{eq:gtphyc}\\
 \phi_d(r) & =  \pm \frac{g_V}{M}\phi_{111}(r). \label{eq:gtphyd}
\end{align}
In LOB,  
$\phi_a$ for $(\kappa_e,\kappa_\nu)=(-1,-1)$ and $(1,1)$ is non zero and is given by
\begin{align}
  \phi_a(r) & = - \sqrt{2} g_A \alpha_{\kappa_e}. 
\end{align}
The spectrum shape $C(E_e)$ is given as
\begin{align}
 (2J_i+1) F(Z,E_e) C(E_e)  
 &= ( \alpha_{-1}^2     +  \alpha_{1}^2) |\braket{f||\sum_{j=1}^A g_A  \tau_j^{\pm} \bm{\sigma}_j||i}|^2 \\
 &= 4\pi ( \alpha_{-1}^2     +  \alpha_{1}^2) 
 \left| g_A \int_0^\infty dr\, r^2 \rho^\sigma_{01}(r)\right|^2.
\end{align}
Note that $\alpha_{-1}^2 + \alpha_1^2$ is often replaced with the
Fermi function evaluated at the nuclear surface~\cite{morita1973beta}.

\subsection{Case for $J^P=0^-$}

For the $J=0$ with the parity change $P=-1$ transition ($0^-$),
only the axial-vector current can contribute. In the impulse approximation, 
the RDME of the multipole operators is
given by the transition densities $\rho^\sigma_{01}$ and $\rho^{\sigma\nabla}_{0}$,
which respectively come from the space $g_A\bm{\sigma}\cdot\bm{r}$
and time $g_A \bm{\sigma}\cdot\bm{\nabla}/M$ components of the axial-vector current.
The latter is velocity dependent.
Their contributions will be equally important as they are estimated by
${\cal O} (\left<p_N\right>/M) \approx 200$ MeV/$1$ GeV
and ${\cal O}(E_e R_A)$, respectively.

The multipole operator for $J^P=0^-$ is given as
\begin{align}
  \braket{f|| \sum_L \Xi_{0L}(\kappa_e,\kappa_\nu)||i}
  & = 
S_{\kappa_e}   \int_0^\infty dr\, r^2\left[
               \rho_{10}^\sigma(r)\phi_b(r)
              + \rho_0^{\sigma\nabla}(r)\phi_c(r)\right],
\end{align}
where $\phi_b$ and $\phi_c$ of Eq. (\ref{eq:ia-unp}) are given as
\begin{align}
 \phi_b(r) & =  -g_A \Phi_{110}(r) - \frac{1}{2M}\frac{d}{dr}
 \left[g_A\phi_{000}(r)+ g_P m_e\bar{\phi}_{000}(r)\right], \label{eq:0mphib}\\
 \phi_c(r) & =  \frac{g_A}{M}\phi_{000}(r). \label{eq:0mphic}
\end{align}

In LOB, two partial waves with $(\kappa_e,\kappa_\nu)=(-1,1),(1,-1)$
can contribute. Then $\phi_b$ and $\phi_c$ are expressed as
\begin{align}
  \phi_b(r) & =  - \sqrt{2}S_{\kappa_e}\alpha_{\kappa_e} 
  \frac{g_A}{3}\left(E_0 + \frac{3V_{D1}(r)}{r} + S_{\kappa_e}m_e\right)r, \\
  \phi_c(r) & =  - \sqrt{2}S_{\kappa_e}\alpha_{\kappa_e} \frac{g_A}{M},
\end{align}
with $V_{D1}(r)$ defined in Eq. (42) of Ref. \cite{horiuchi2021electron},
\begin{align}
  V_{D1}(r)  & =  - \int_0^r  \left(\frac{r'}{r}\right)^{2k} V_C(r') dr'.
\end{align}
In order to compare with LOB for the first-forbidden transition,
we rewrite the RDMEs using the parametrization of Ref. \cite{sch66}
summarized in the appendix. The RDME in LOB is written as
\begin{align}
  \sqrt{2\pi}\braket{f||  \sum_L \Xi_{0L}(\kappa_e,\kappa_\nu)||i}
  & =  - \alpha_{\kappa_e} \left( \zeta_0 + S_{\kappa_e}\frac{m_e w}{3}\right).
\end{align}
Here $\zeta_0$ and $w$ are the nuclear matrix elements defined in the appendix.
The spectrum shape is given as
\begin{align}
  (2J_i+1)F(Z,E_e) C(E_e) & = 
      \alpha_{-1}^2 \left(\zeta_0 - \frac{m_e w}{3}\right)^2
    +  \alpha_{1}^2 \left(\zeta_0 + \frac{m_e w}{3}\right)^2,
\end{align}
which is in agreement with Ref. \cite{sch66}. 
We note that this formula is often used
with the following approximations
\begin{align}
  \alpha_{-1}^2 + \alpha_1^2 & \sim  F(Z,E_e), \label{eq:approx1}\\
  \lambda_2, \mu_1 & \sim  1. \label{eq:approx2}
\end{align}
The accuracy of the above treatment depends on $Z$ and $E_e$.
Deviation from the approximations of Eqs. (\ref{eq:approx1})
and (\ref{eq:approx2})
can be very large for large $Z$~\cite{horiuchi2021electron,behrens1969numerical}.


\subsection{Case for $J^P=1^-$ }

The operators for the $J=1$ and parity $P=-1$ transition include
the contribution of $V_0$, $\bm{V}$, and $\bm{A}$.
The multipole operator in the impulse approximation is given by
\begin{align}
  \braket{f|| \sum_L \Xi_{1L}(\kappa_e,\kappa_\nu)||i}
  & = 
S_{\kappa_e}   \int_0^\infty dr\, r^2\left[
               \rho_{1}(r)\phi_A(r)
              + \rho_{11}^\sigma(r)\phi_B(r)
              + \rho_{21}^{\nabla}(r)\phi_C(r)\right]
              \label{eq:1-}
\end{align}
with
\begin{align}
  \phi_A(r) & = \mp g_V\left[\Phi_{011}(r) + \sqrt{3}\omega \frac{\varphi_1(r)}{r^2} - \frac{\sqrt{2}}{2M}\psi_1(r)\right],\label{eq:1mphia}
\\
  \phi_B(r) & = - g_A \Phi_{111}(r)
  \mp \frac{g_V+g_M}{2M}\psi_1(r), \label{eq:1mphib}
  \\
  \phi_C(r) & = \pm \frac{g_V}{M}\left\{\phi_{121}(r) +
  \sqrt{2}\left[\phi_{101}(r)-3\frac{\varphi_1(r)}{r^3}\right]\right\}, \label{eq:1mphic}
\end{align}
where
\begin{align}
  \psi_1(r) & =  \sqrt{\frac{2}{3}}\frac{d}{dr}\phi_{101}(r)
  +\sqrt{\frac{1}{3}}\left(\frac{d}{dr}+\frac{3}{r}\right)\phi_{121}(r).
\end{align}
When we evaluate the convention current without using the current conservation
relation,  the multipole operator $\Xi$ is given as
\begin{align}
  \braket{f|| \sum_L \Xi_{1L}(\kappa_e,\kappa_\nu)||i}
  & = 
S_{\kappa_e}   \int_0^\infty dr\, r^2 \notag\\
&\times \left[
               \rho_{1}(r)\phi_{A'}(r)
              + \rho_{11}^\sigma(r)\phi_B(r)
              + \rho_{21}^{\nabla}(r)\phi_{C'}(r)
              + \rho_{01}^{\nabla}(r)\phi_{D}(r)
\right], \label{eq:no_cur}
\end{align}
where the leading-order convection current term $\rho_{01}^\nabla$ 
is given together with the lepton wave function
\begin{align}
  \phi_{D}(r) & = \pm \frac{g_V}{M}\phi_{101}(r),
\end{align}
 and $\phi_{A'}$ and $\phi_{C'}$ are:
 \begin{align}
   \phi_{A'}(r) & = \mp g_V[  \Phi_{011}(r) -\frac{1}{2M}\psi_1'(r)] , \\
  \phi_{C'}(r) & = \pm \frac{g_V}{M}\phi_{121}(r),
 \end{align}
 where
 \begin{align}
  \psi_1'(r) & =  \sqrt{\frac{1}{3}}\frac{d}{dr}\phi_{101}(r)
  - \sqrt{\frac{2}{3}}\left(\frac{d}{dr}+\frac{3}{r}\right)\phi_{121}(r).
\end{align}

 In LOB, lepton wave functions are approximated as explained in I
 and the contributions of ${\cal O}(1/M)$ terms are neglected.
 The RDME is given by the dipole $\rho_1$ and spin-dipole $\rho_{11}^\sigma$ terms
 as
\begin{align}
  \braket{f|| \sum_L \Xi_{1L}(\kappa_e,\kappa_\nu)||i}
  & = 
S_{\kappa_e}   \int_0^\infty dr\, r^2 \left[
               \rho_{1}(r)\phi_A(r)
              + \rho_{11}^\sigma(r)\phi_B(r)\right].
\end{align}
Here, $\rho_{1}\phi_A$ and $\rho^{\sigma}_{11}\phi_B$ terms
correspond to $g_V \bm{r}$ and $g_A \bm{r} \times \bm{\sigma} $, respectively.
Six partial waves with
$(\kappa_e,\kappa_\nu)=(\mp 1,\pm 1),(\mp 1, \mp 2),(\mp 2,\mp 1)$ contribute. 
For the partial waves with $(\kappa_e,\kappa_\nu)=(-1,-2),(1,2),(-2,-1)$, and $(2,1)$,
$\phi_A$ and $\phi_B$ are given by
\begin{align}
  \phi_A(r) & = \mp g_V \alpha_{\kappa_e}\frac{2\sqrt{3}}{9}
  \left(p_\nu \delta_{|\kappa_e|,1} + p_e \delta_{|\kappa_e|,2}\right)r, \\
  \phi_B(r) & =    g_A \alpha_{\kappa_e}\frac{\sqrt{6}}{9}
  \left(p_\nu \delta_{|\kappa_e|,1} - p_e \delta_{|\kappa_e|,2}\right)r.
\end{align}
The RDME can be written as
\begin{align}
  \sqrt{2\pi}\braket{f||  \sum_L \Xi_{1L}(\kappa_e,\kappa_\nu)||i}
  & =  S_{\kappa_e} \alpha_{\kappa_e} \frac{1}{3\sqrt{2}}
 \left[ (2x + u) p_\nu\delta_{|\kappa_e|,1} + (2x - u) p_e\delta_{|\kappa_e|,2}\right].
\end{align}
Here, $x$ and $u$ are defined in the appendix.

For the partial waves with $(\kappa_e,\kappa_\nu)=(-1,1)$ and $(1,-1)$,
$\phi_A$ and $\phi_B$ are given as
\begin{align}
  \phi_A(r) & =  \pm g_V S_{\kappa_e}\alpha_{\kappa_e}\frac{\sqrt{6}}{9}
  \left(E_0 + \frac{3V_{D1}}{r} + 3\omega + S_{\kappa_e}m_e\right)r, \\
  \phi_B(r) & =  g_A S_{\kappa_e}\alpha_{\kappa_e}\frac{2\sqrt{3}}{9}
  \left(E_0 - 2E_e - \frac{3V_{D1}}{r} - S_{\kappa_e}m_e\right)r, 
\end{align}
and
\begin{align}
  \sqrt{2\pi}\braket{f||  \sum_L \Xi_{1L}(\kappa_e,\kappa_\nu)||i}
  & =  \alpha_{\kappa_e} 
 \left[\zeta_1 - \frac{2}{3}u E_e - S_{\kappa_e}\frac{m_e}{3}(x+u)\right].
\end{align}
From the Siegert's theorem,  the RDME of
the convection current is written in terms of that of the dipole operator;
the following relation is used in the above formula
\begin{eqnarray}
  \xi' y & \rightarrow & - \omega x.
\end{eqnarray}  
Here $\zeta_1$ and $\xi'y$ are defined in the appendix.

Finally, the spectrum shape is given as 
\begin{align}
 (2J_i+1) F(E_e,Z)C(E) & =  (\alpha_{-1}^2 + \alpha_1^2)\left[
    \frac{(2x+u)^2 p_\nu^2 }{18}
   +   \lambda_2 \frac{(2x-u)^2 p_e^2}{18}\right]
   \notag \\
   & 
   + \sum_{\kappa_e=\mp 1} \alpha_{\kappa_e}^2
        \left[\zeta_1 - \frac{2}{3}u E_e - S_{\kappa_e} \frac{m_e}{3}(x+u)\right]^2.
\end{align}
The above expression agrees with
that given in Ref.~\cite{sch66}.


\subsection{Case for $J^P=2^-$}

The $2^-$ transition offers a unique first-forbidden transition, where
only the single spin-dipole operator $[Y_1\otimes \bm{A}]_2$($\rho_{12}^\sigma$)
contributes in LOB.
Due to this property, the $2^-$ transition has been used 
for a precise evaluation of the beta-ray spectrum in 
BSM searches.
However, by including NLO, we have an additional contribution 
of the weak magnetism, $[Y_3\otimes \bm{A}]_2$($\rho_{32}^\sigma$),
the time component of the axial-vector current $Y_2 A_0$($\rho_2^{\sigma\nabla}$),
and the vector current $[Y_2\otimes \bm{V}]_2$($\rho_{22}^\nabla$).
In the impulse approximation, the multipole operator is given by
\begin{align}
  \braket{f|| \sum_L \Xi_{2L}(\kappa_e,\kappa_\nu)||i}
  & = 
S_{\kappa_e}   \int_0^\infty dr\, r^2 \notag\\ 
&\times \left[ \rho_{12}^\sigma(r)\phi_a(r)
               + \rho_{32}^\sigma(r)\phi_b(r)
               + \rho_2^{\sigma\nabla}(r)\phi_c(r)
               +  \rho_{22}^\nabla(r)\phi_d(r)\right] \label{eq:2-}
\end{align}
with
\begin{align}
  \phi_a(r) & = 
  - g_A \Phi_{1 1 2}(r)
 +  \frac{1}{2M}\sqrt{\frac{2}{5}}D_-^2 \left(g_A\phi_{022}(r)+g_P m_e\bar{\phi}_{022}(r)\right)\notag\\
& \mp \frac{g_V + g_M}{2M}\sqrt{\frac{3}{5}}D_-^2 \phi_{122}(r), \\
  \phi_b(r) & = 
  - g_A \Phi_{1 3 2}(r)
 -  \frac{1}{2M}\sqrt{\frac{3}{5}}D_+^2 \left(g_A \phi_{022}(r) + g_P m_e \bar{\phi}_{022}(r)\right)\notag\\
 &\mp \frac{g_V + g_M}{2M}\sqrt{\frac{2}{5}}D_+^2 \phi_{122}(r), \\
 \phi_c(r) & =  \frac{g_A}{M}\phi_{022}(r), \\
 \phi_d(r) & =  \pm \frac{g_V}{M}\phi_{122}(r).
\end{align} 

For $|\kappa_e|,|\kappa_\nu| \le 2$, six partial waves
with $(\kappa_e,\kappa_\nu)$ $=(\pm 1,\pm 2),(\pm 2, \pm 1),(\pm 2, \mp 2)$
contribute.
In LOB, where only the operator $[Y_1 \otimes \bm{\sigma}]_2$  contributes,
the RDME for four partial waves 
with $(\kappa_e,\kappa_\nu)=(\pm 1,\pm 2),(\pm 2, \pm 1)$ 
is given as
\begin{align}
  \braket{f|| \sum_L \Xi_{2L}(\kappa_e,\kappa_\nu)||i}
  & = 
S_{\kappa_e}   \int_0^\infty dr\, r^2\rho_{12}^\sigma(r)\phi_a(r)
\end{align}
with
\begin{align}
  \phi_a(r) & =  - g_A \alpha_{\kappa_e}\frac{\sqrt{2}}{3}
  \left(p_\nu \delta_{|\kappa_e|,1} + p_e \delta_{|\kappa_e|,2}\right) r.
\end{align}

By using the parametrization of the matrix elements given in the appendix, 
the RDME reads 
\begin{align}
  \sqrt{2\pi}\braket{f||  \sum_L \Xi_{JL}(\kappa_e,\kappa_\nu)||i}
  & = S_{\kappa_e} \alpha_{\kappa_e} \frac{z}{2\sqrt{3}}
 \left(  p_\nu\delta_{|\kappa_e|,1} + p_e\delta_{|\kappa_e|,2}\right),
\end{align}
and the spectrum shape is given by
\begin{align}
(2J_i+1) F(E_e,Z) C(E_e) & =  \left(\alpha_{-1}^2 + \alpha_1^2\right)\frac{z^2}{12} \left( p_\nu^2 + 
 \lambda_2 p_e^2\right).
\end{align}
The expression agrees with that obtained in Ref.~\cite{sch66}.

\subsection{Beta-decay rate of $^{160}$Sn}

\begin{table}[h]  
\begin{center}
\caption{\label{tab:all}
Beta-decay rate in sec$^{-1}$. 
Electron wave functions obtained by solving the Dirac equation are employed for `Exact', while the approximated electron wave functions 
described in paper (I) are used for NLO and LOB.
}
\begin{tabular}{cc|cc|ccc|c}\hline
  $J^P$    &        &  $\bm{A}$ &   $+V_0$  & $+A_0$  & $+\bm{V}(\rm{WM})$ & $+\bm{V}(\rm{Conv.})$ & Total \\ \hline
  $1^+$ & Exact  &  280.     &   -    &  285. &  296.       & 296. & 296.    \\
        & NLO    &  279.     &   -    &  285. &  295.       & 295. & 295.    \\
        & LOB    &  320.    &   -     &  -      &  -            & -      & 320.       \\ \hline
  $0^-$ & Exact    &   51.4    &  -      &  14.1  &  -           &   -    & 14.1    \\
        &  NLO    &  51.2      &   -     &  14.1  &   -         &    -    & 14.1     \\
        & LOB     &    60.3      & -       &  14.1 &  -           &   -    & 14.1    \\     \hline
  $1^-$ & Exact  &  40.1         & 75.8      & -          & 78.3     & 90.0 & 90.0   \\
        & NLO    &  40.0         & 75.5      & -          & 78.2     & 90.3 & 90.3 \\
        & LOB    &  45.5         & 87.5      & -          & -         & 93.4 & 93.4   \\ \hline
  $2^-$ & Exact  &  24.3        & -    & 25.1      &  26.2    & 26.6 & 26.6   \\
        & NLO    &  24.2        & -   &  25.0      & 26.1     & 26.5 & 26.5 \\
        & LOB   &   28.7        & -   & -          &  -        & -     & 28.7   \\
\hline
\end{tabular}
\end{center}
\end{table}

The formulas of the beta-decay rate in the impulse approximation developed in the
previous subsection are applied to $^{160}$Sn.
The total decay rate is obtained by the sum of the partial decay rates of about 1400
states obtained from the pnQRPA approach.
The allowed transition to the $1^+$ states and the first-forbidden transitions
to the $0^-,1^-,$ and $2^-$ states are examined.
We take $g_{A}=1$ and include partial waves
of the lepton wave function up to $|\kappa_e|,|\kappa_\nu|  \le2$.

The results are summarized in Tab.~\ref{tab:all}. 
By `Exact', we use the electron wave function obtained from the numerical solution
of the Dirac equation and the plane-wave neutrino wave function.
An approximate electron wave function of NLO is described in paper (I).
The decay rate calculated from the LO terms of the nucleon current
in the non-relativistic approximation is given in the third (the space component of the axial-vector current, $\bm{A}$) and
the fourth column (adding the time component of the vector current, $V_0$).
The 5th, 6th, and 7th columns show the decay rates calculated from the
$\bm{A}+V_0+A_0$, $\bm{A}+V_0+A_0+\bm{V}(\rm{WM})$,
and $\bm{A}+V_0+A_0+\bm{V}(\rm{WM})+\mbox{\boldmath $V$}(\rm{Conv.})$ terms, respectively.
The decay rates including all the components of the nucleon current
are given in the 8th column (Total). Here the `Convection current' (Conv.) includes
the term obtained using the current conservation relation
and all the terms associated with the convection current 
as explained in the previous subsection.
In all the cases, NLO reproduces the `Exact' results with 1\% accuracy.

For the allowed transition $1^+$ and the unique first-forbidden transition $2^-$,
the $\bm{A}$ term gives a dominant contribution.
The $A_0$, $\bm{V}(\rm{WM})$, and $\bm{V}(\rm{Conv.})$ terms respectively contribute to the
decay rate approximately by 2\%, 3\% and $-$0.1\% for $1^+$
and 3\%, 4\% and 2\% for $2^-$.
LOB overestimates the decay rate by 14\% for $1^+$ and 12\% for $2^-$
when `Exact' is calculated using the $\bm{A}$ term alone, while
 it reduces to 8\% for $1^+$ and 7\% for $2^-$ 
 when the $A_0$ and $\bm{V}(\rm{WM})$ terms are included in `Exact'.
The contributions of higher-order terms $Y_2 \bm{\sigma}\cdot\nabla$
and $[Y_2 \otimes \nabla]_2$ are both less than 1\% and their contributions
are negligible.

\begin{figure}
\begin{center}  
  \includegraphics[width=10cm]{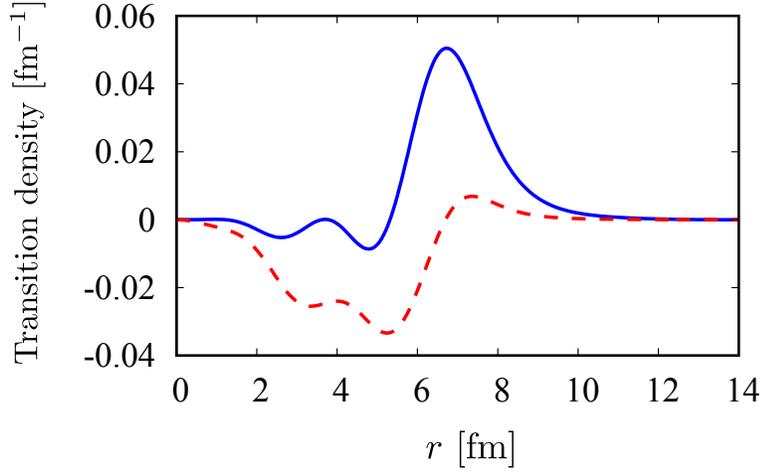}
  \caption{\label{fig:0-trdn}
  Transition densities 
   for the $0^-$ state at $E_0=13.94$ MeV of $^{160}$Sn. 
  The solid (blue) and dashed (red) curves show
  $\rho^\sigma_{10} r^3 E_0$ and $(\rho^{\sigma \nabla}_0/M) r^2$,
  respectively.
  }
\end{center}
\end{figure}

The ${\cal O}(p/M)$ currents play an important role in the $0^-$ and $1^-$ transitions.
For the transition to the $0^-$ states,
the velocity-dependent current $A_0$ is as important as
the spin-dipole operator in the $\bm{A}$ term. 
Tab. \ref{tab:01-} shows the decay rate calculated
using the $\bm{A}$, $V_0$, $A_0$, and $\bm{V}(\rm{Conv.})$ terms for $0^-$ and $1^-$.
A strong cancellation between two components takes place.
This can be seen from the transition densities shown 
in Fig.~\ref{fig:0-trdn} for the state 
with the largest partial decay width.
Although the effect of improving the treatment of lepton
wave function from LOB to `Exact' is about 
17\% for the $\bm{A}$ term and 9\% for the $A_0$ term,
the effect becomes small for the total decay rate in this case.
Since the pion-exchange current due to the soft-pion production is known to contribute as largely as the impulse current~\cite{kdr},
it would be important to include the meson-exchange current
for the realistic estimation of the decay rate of $0^-$~\cite{kdr,towner1986}.

\begin{table}[h]
\begin{center}
\caption{\label{tab:01-} 
Calculated beta-decay rates in sec$^{-1}$ 
for the $\bm{A}$, $V_0$, $A_0$ and $\bm{V}(\rm{Conv.})$ terms.
}
\begin{tabular}{c|cc|cc}\hline
$J^P$      & $\bm{A}$ & $V_0$   & $A_0$ & $\bm{V}(\rm{Conv.})$ \\ \hline
$0^-$ &   51.4   &    -    & 110. &   -    \\
$1^-$ &  40.1    &   64.9  &    -  &  114. \\
\hline
\end{tabular}
\end{center}
\end{table}

\begin{figure}
\begin{center}
  \includegraphics[width=10cm]{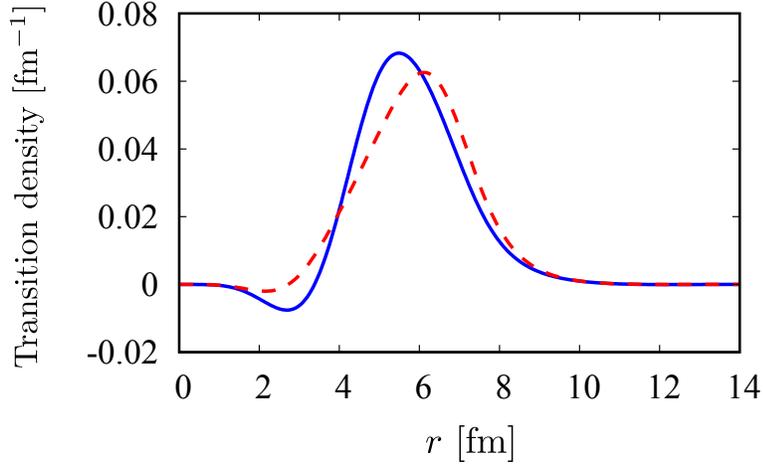}
  \caption{Transition densities 
  for the $1^-$ state at $E_0=15.63$ MeV of $^{160}$Sn.
The solid (blue) and dashed (red) curves show
$\rho_1 r^3 E_0$ and $\rho^\sigma_{11}r^3 E_0$, respectively.
  } 
\label{fig:1-trd}
\end{center}
\end{figure}

For the transition to the $1^-$ states,
the dipole operator from the $V_0$ and $\bm{V}(\rm{Conv.})$ terms
and the spin-dipole operator from the $\bm{A}$ term are equally important. 
Their contributions are shown in Tab. \ref{tab:01-}.
If we do not use the current conservation relation for the
convection current  but use Eq. (\ref{eq:no_cur}),
the calculated decay rate becomes 140 sec$^{-1}$.
The use of the current conservation relation
for the electric multipole of the vector current reduces 
the total decay rate by about 45\%,  
while LOB overestimates the decay rate by about 5\%. 
The weak magnetism alters the decay rate by about 5\%, 
which is sizable, while
the contribution of the $\rho_{21}^\nabla$ term is less than 1\%.
The transition densities for the dipole 
and spin-dipole operators are shown in Fig. \ref{fig:1-trd}. 
One sees coherent contributions of the $V_0$ and $\bm{A}$ terms.

\section{Summary}
\label{sec:summary}

The formulas of the beta-decay rate including the velocity-dependent weak currents as well as the induced currents have been
presented for the allowed and first-forbidden transitions.
The longitudinal part of the vector current is eliminated 
by using the conservation of the vector current 
for the multipole operator of $P=(-1)^J$ transition.

The allowed and first-forbidden transitions of the beta decay of $^{160}$Sn
have been studied as an application of the present formulas.
The transition densities are obtained by the EDF method.
The results obtained with the electron wave functions in LO and NLO are compared with the `exact' ones.
It is shown that the conventional LO formula misses the transition rate
by about 10 to 20 $\%$.
Including the NLO term reproduces the `exact' with high accuracy.
The induced weak-magnetism term typically contributes by about a few percent
to the decay rate through the interference term  of the leading-order amplitude. 
Reorganizing the formula using the current conservation is important for the $1^-$ transition operator of the vector current as known well. 
We reconfirmed the velocity-dependent term is as equally important as
the spin-dipole term for the $0^-$ transition. 
The induced pseudo-scalar term can be neglected for the nuclear beta decay.

The formulas offer semi-analytic expressions of the
transition matrix element of the nuclear beta decay,
which are transparent and allow us to apply not only to various beta-decay processes but also the shape and angular correlations of the beta decay and neutrino reactions. The precise formulas  are also
useful to search signals of BSM in nuclear beta decay.


\section*{Acknowledgment}
We would like to thank Prof. K. Koshigiri for the useful discussions.
This work was in part supported by the JSPS KAKENHI
Grants Nos. JP18H01210,  JP18H04569,  JP18K03635,
JP19H05140, JP19K03824, JP21H00081, JP22H01237, JP22K03602, and JP22H01214,
 the Collaborative Research Program 2022, Information Initiative Center, Hokkaido University, 
and the JSPS/NRF/NSFC A3 Foresight Program ``Nuclear Physics in the 21st Century.'' 
The nuclear EDF calculation was performed on Yukawa-21 
at the Yukawa Institute for Theoretical Physics, Kyoto University.


\appendix

\section{Nuclear matrix element}

Nuclear matrix elements $w,x,u,z$ and those with Coulomb potential and velocity-dependent operators that are used in the literature are expressed in terms of the
transition densities introduced in section \ref{sec:applications}:
\begin{align}
  w & =  g_A \sqrt{4\pi}  \int_0^\infty dr r^3 \rho^\sigma_{10}(r), \\
  x & = \mp g_V  \sqrt{\frac{4\pi}{3}} \int_0^\infty dr r^3 \rho_1(r), \\
  u & = g_A \sqrt{\frac{8\pi}{3}}   \int_0^\infty dr r^3 \rho^\sigma_{11}(r), \\
  z & = - g_A \sqrt{\frac{16\pi}{3}}  \int_0^\infty dr r^3 \rho^\sigma_{12}(r), \\
    \xi  w' & = g_A  \sqrt{4\pi} \int_0^\infty dr r^2 \rho^\sigma_{10}(r)V_{D1}(r), \\
\xi x' & = \mp g_V \sqrt{\frac{4\pi}{3}} \int_0^\infty dr r^2 \rho_1(r)V_{D1}(r),\\
  \xi  u' & =  g_A\sqrt{\frac{8\pi}{3}}  \int_0^\infty dr r^2 \rho^\sigma_{11}(r)
    V_{D1}(r), \\
    \xi'v &= \sqrt{4\pi}\frac{g_A}{M}\int_0^\infty dr r^2 \rho^{\sigma\nabla}_0(r),\\
    \xi'y &= \mp \sqrt{4\pi}\frac{g_V}{M}\int_0^\infty dr r^2 \rho^{\nabla}_{01}(r),\\
    \zeta_0 & = \xi w' + \xi' v + \frac{E_0 w}{3}, \\
    \zeta_1 & = \xi' y - \xi u' - \xi x' + \frac{(u - x)E_0}{3}
\end{align}
for $\beta^\mp$ decay and
\begin{align}
  \lambda_2 & = \frac{\alpha_{-2}^2 + \alpha_{2}^2}{\alpha_{-1}^2 + \alpha_{1}^2}, \\
  \frac{m_e}{E_e}\gamma_1\mu_1 & = 
  \frac{\alpha_{-1}^2 - \alpha_{1}^2}{\alpha_{-1}^2 + \alpha_{1}^2}.
\end{align}



\bibliographystyle{ptephy}
\bibliography{beta_decay}

\end{document}